# Magnetic damping modulation in IrMn$_3$/Ni$_{80}$Fe$_{20}$ via the magnetic spin Hall effect

José Holanda[1*], Hilal Saglam[1,2], Vedat Karakas[3], Zhizhi Zang[1,4], Yi Li[1], Ralu Divan[5], Yuzi Liu[5], Ozhan Ozatay[3], Valentine Novosad[1], John E. Pearson[1] and Axel Hoffmann[1,6*]

[1]Materials Science Division, Argonne National Laboratory, Lemont, Illinois 60439, USA
[2]Illinois Institute of Technology, Chicago, Illinois 60616, USA
[3]Physics Department, Bogazici University, North Campus, Bebek/Istanbul 34342, Turkey
[4]School of Optical and Electronic Information, Huazhong University of Science and Technology, Wuhan 430074, People's Republic of China
[5]Center for Nanoscale Materials, Argonne National Laboratory, Argonne, IL 60439, USA
[6]Department of Materials Science and Engineering, University of Illinois at Urbana-Champaign, Urbana, IL 61801, USA

Non-collinear antiferromagnets can have additional spin Hall effects due to the net chirality of their magnetic spin structure, which provides for more complex spin-transport phenomena compared to ordinary non-magnetic materials. Here we investigated how ferromagnetic resonance of permalloy (Ni$_{80}$Fe$_{20}$) is modulated by spin Hall effects in adjacent epitaxial IrMn$_3$ films. We observe a large dc modulation of the ferromagnetic resonance linewidth for currents applied along the [001] IrMn$_3$ direction. This very strong angular dependence of spin-orbit torques from dc currents through the bilayers can be explained by the magnetic spin Hall effect where IrMn$_3$ provides novel pathways for modulating magnetization dynamics electrically.

*Corresponding author: joseholanda.papers@gmail.com

Antiferromagnetic materials are promising for future spintronic applications, since they combine several advantageous features. They are robust against perturbation due to magnetic fields, produce no stray fields, display ultrafast dynamics, and are capable of generating large magnetotransport effects [1, 2]. The idea of using antiferromagnetic materials in spintronic devices [3, 4] has gained interest with the realization that antiferromagnets can be efficient sources of spin currents [5-9] and that their spin structure can be modulated electrically [10-12]. Furthermore, antiferromagnets with non-collinear spin configurations provide additional rich new spin-transport phenomena, since any chirality of their spin structure may result in non-vanishing Berry curvatures affecting profoundly their charge transport properties [13-15]. Towards this end, it has been shown that triangular antiferromagnets with chiral spin arrangements can exhibit ferromagnetic-like behaviors such as a large anomalous Hall and Nernst effects, as well as a magneto-optical Kerr effect. An interesting addition to these discoveries is the magnetic spin Hall effect (MSHE), which has recently been observed in the triangular antiferromagnet $SnMn_3$ [16]. The magnetic spin Hall effect originates from a reactive counterpart of the dissipative spin response responsible for the ordinary spin Hall effect [13]. This interpretation is supported by the dependence of the spin Hall effect (SHE) signals on the magnetic-order parameter reversal and can be interpreted in terms of the symmetries of well-defined linear response functions. This means that the MSHE can be explored under the condition of ferromagnetic resonance and represents a new way of analyzing the magnetization dynamics, *i.e.*, when *dc* electrical current passes through the uniformly magnetized material, a non-equilibrium distribution of spins at the interface influences the dynamic properties. In particular, current-induced modulation of damping has become a standard technique for quantifying spin Hall effects [17, 18]. Therefore the control of damping in ferromagnetic/antiferromagnetic bilayer can provide fundamental insights for ferromagnetic or antiferromagnetic spintronics.

      Here we investigate the non-collinear antiferromagnet $IrMn_3$, in which Mn atoms form Kagome lattice in the {111} planes [See **Fig. 1 (a)**]. In addition, $IrMn_3$ has a giant magnetocrystalline anisotropy energy due to the locally broken cubic symmetry of the Mn sublattices [19]. Additionally, $IrMn_3$ has a large variation of the anomalous Hall conductivities along different crystallographic orientations, *i.e.*, $\sigma_{[011]} = 224 \ \Omega^{-1}cm^{-1}$ and $\sigma_{[001]} = 8.8\times10^3 \ \Omega^{-1}cm^{-1}$ for [011] and [001] directions, respectively [20-22]. This represents a

ratio $\sigma_{[011]}:\sigma_{[001]} = 1:39$ between the spin Hall conductivities. In this letter, we perform ferromagnetic resonance measurements with and without *dc* current in IrMn$_3$/Ni$_{80}$Fe$_{20}$ bilayers. Our results show a strong angular dependence of the electric current induced modulation of ferromagnetic resonance linewidth for magnetic fields applied along different crystalline orientations, *i.e.*, [001] and [011], while the *dc* current is applied along the [001] direction. We report a maximum Gilbert damping modulation of 41% and the observed anisotropy for the magnetic fields applied along the two different crystalline directions can be associated with the variation of spin accumulation at the interface due to the additional contributions arising from the magnetic spin Hall effect.

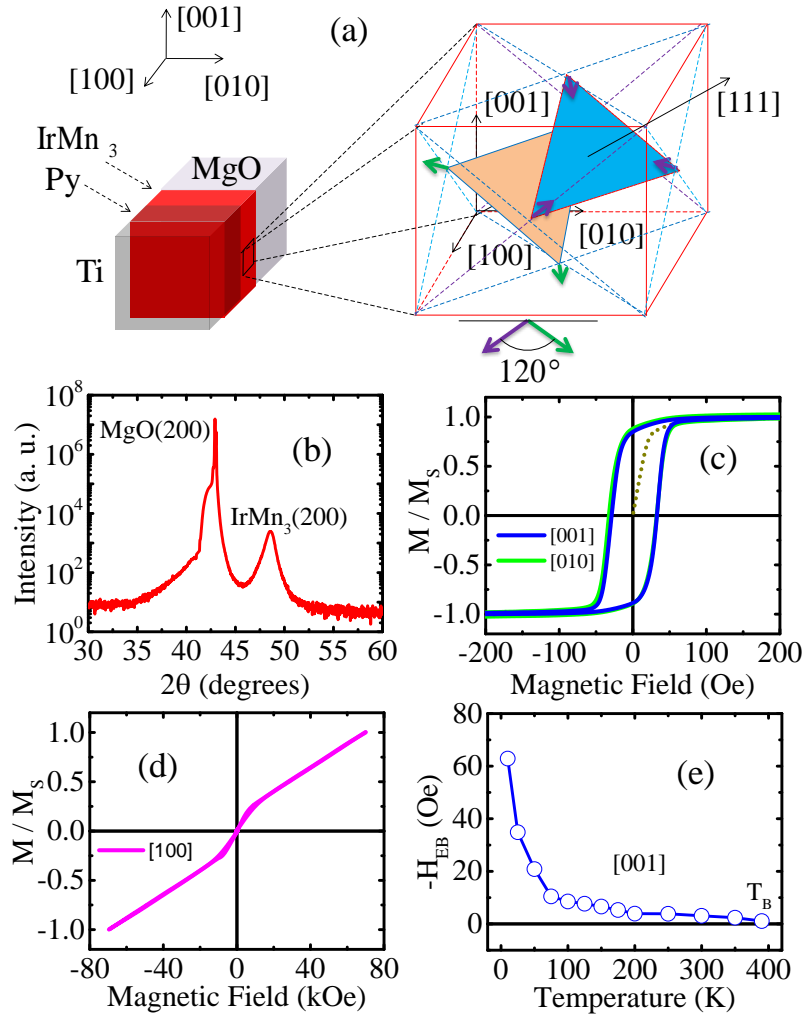

**Figure 1: (a)** shows the schematic of IrMn$_3$/Ni$_{80}$Fe$_{20}$ (permalloy, Py) bilayer capped with a thin layer of Ti and also includes the unit cell of IrMn$_3$ with its spin structure. Note that the Mn atoms

in the unit cell of IrMn$_3$ form a Kagome lattice in the {111} planes with the spins either pointing towards or away from each triangular Mn arrangement. **(b)** The X-ray diffraction pattern measured for a 20-nm thick IrMn$_3$ layer capped with a 2-nm Ti layer. **(c)** Measured magnetic hysteresis curves after field-cooling in a magnetic field of +70 kOe applied along either [001] or [010] directions at a temperature of 300 K. **(d)** Magnetic hysteresis curves after field-cooling in a magnetic field of +70 kOe applied along the [100] direction at temperature of 300 K. **(e)** Measured exchange bias field as a function of temperature with applied magnetic field of +70 kOe along the [001] direction, showing that the blocking temperature of IrMn$_3$(20 nm)/Ni$_{80}$Fe$_{20}$(10 nm) bilayer is 380 K.

We have grown nominally 20-nm thick epitaxial IrMn$_3$ films on MgO (100)-oriented single-crystal substrates at 570 $^o$C using a magnetron sputtering technique. Subsequently, a thin layer of Ni$_{80}$Fe$_{20}$ (10 nm) and Ti(2 nm) were deposited at room temperature. Here, the Ti layer was used to protect the surface properties of Ni$_{80}$Fe$_{20}$. **Figure 1 (a)** shows the unit cell of IrMn$_3$, where the Mn moments are parallel to the {111} planes and aligned along the <112> directions. **Figure 1 (b)** shows an X-ray diffraction pattern for IrMn$_3$ grown on a MgO substrate. We found that IrMn$_3$ has a lattice constant of (0.377 ± 0.001) nm, which is consistent with previous literature values [20, 23].

The magnetic characterization was performed with a Superconducting Quantum Interference Device (SQUID) magnetometer. **Figures 1 (c)**, **(d)** and **(e)** are measured for MgO/IrMn$_3$(20 nm)/Ni$_{80}$Fe$_{20}$(10 nm)/Ti(2 nm). In **Figures 1 (c)** and **(d)** we show the hysteresis curves at room temperature (300 K) after field-cooling starting in temperature of 390 K with a magnetic field of +70 kOe applied along the crystallographic directions [001], [010] and [100]. **Figure 1 (e)** shows the exchange bias field as a function of temperature for magnetic fields applied along the [001] direction, where exchange bias is defined as H$_{EX}$ = - (H$_1$ + H$_2$)/2. This means that the blocking temperature is above 300 K, which shows that an exchange bias can be stabilized at room temperature for IrMn$_3$/Py bilayer and thus IrMn$_3$ is antiferromagnetically ordered at room temperature [23].

We utilize a flip-chip ferromagnetic resonance technique [24] for characterizing the Gilbert damping measured for fields applied in different crystal orientations. We measured the transmission coefficient by sweeping the frequency at fixed fields. **Figure 2 (a)** shows the experimental configuration for ferromagnetic resonance measurements with and without *dc* currents. **Figures 2 (b)** and **(e)** show the ferromagnetic resonance (FMR) signals obtained with a vector network analyzer (VNA) for magnetic fields applied along the [001] and [011] crystallographic directions, respectively.

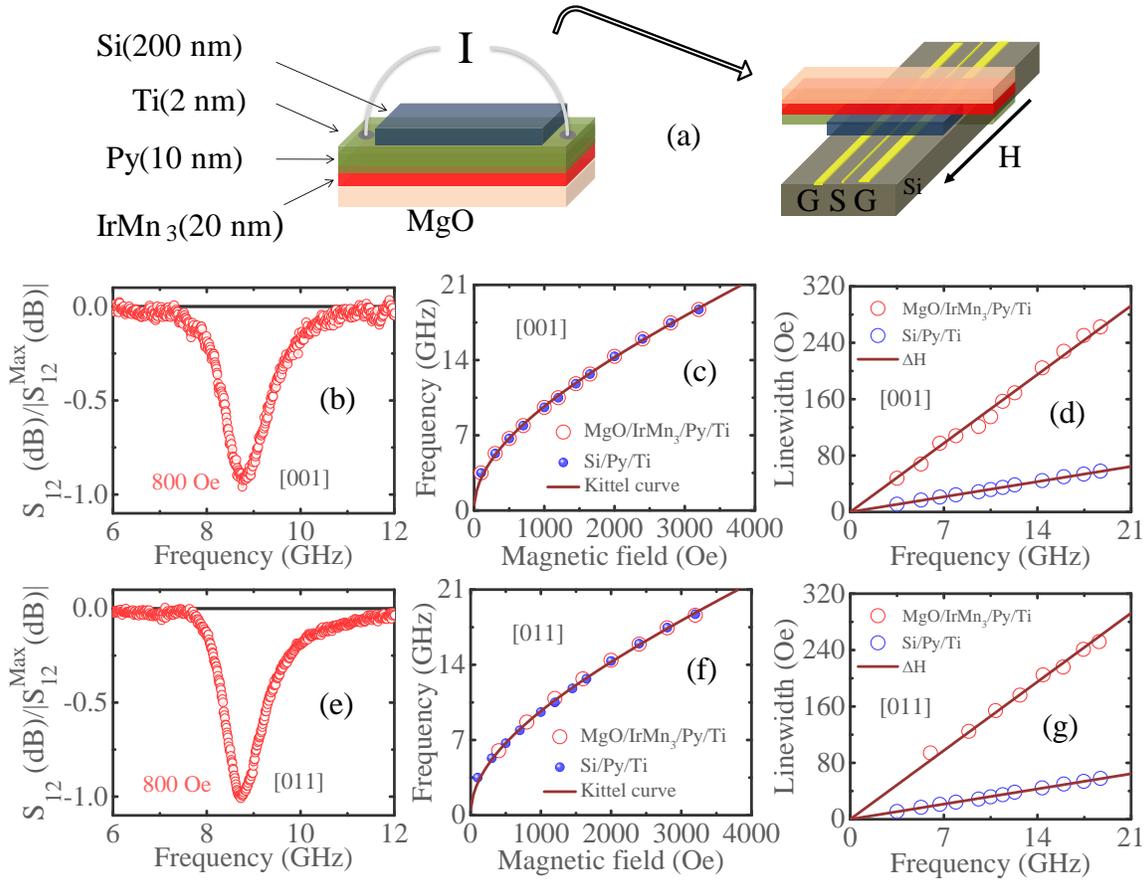

**Figure 2: (a)** Schematic of the flip-chip ferromagnetic resonance measurement set up, where *dc* current (I) leads are connected to the Ti layer. **(b)** and **(e)** ferromagnetic resonance (FMR) signals obtained using a VNA for the magnetic field applied along the [001] and [011] crystallographic directions, respectively. **(c)** and **(f)** FMR frequency as a function of magnetic field, which was applied along the [001] and [011] crystallographic directions, respectively. The fits are performed with the Kittel equation, where $\gamma = g\mu_B/\hbar = 2.8$ GHz/kOe and $4\pi M_{eff} = (9706 \pm 1)$ Oe. **(d)** and **(g)** show the linewidth variation as a function of the FMR frequency for the crystallographic directions [001] and [011], respectively.

**Figures 2 (c)** and **(f)** show the measured FMR frequency as a function of the magnetic field. The solid curve was obtained by fitting the experimental data to the Kittel equation, $f = \gamma[(H_R)(H_R + 4\pi M_{eff})]^{1/2}$ where $\gamma = g\mu_B/\hbar = 2.8$ GHz/kOe is the gyromagnetic ratio, $g = 2$ is the spectroscopic splitting factor, $\mu_B$ is the Bohr magneton, $\hbar$ the reduced Planck constant, and $4\pi M_{eff}$ the effective magnetization, which is determined by the fit $4\pi M_{eff} = (9706 \pm 1)$ Oe. As expected, there is no significant variation in the ferromagnetic resonance field for single $Ni_{80}Fe_{20}$ or $Ni_{80}Fe_{20}/IrMn_3$ samples. The frequency swept linewidths ($\Delta f_{VNA}$) were obtained via Lorentz fitting. Detailed steps, including the conversion from $\Delta f_{VNA}$ to $\Delta H$ can be found in Ref. 24. Thus the linewidth as a function of the frequency is written by considering equation $\Delta H = \Delta H_0 + (\alpha/\gamma)f$ [25], where $\alpha$ is the magnetic Gilbert damping of the material. **Figures 2 (d)** and **(g)** represent the linewidth considering the transformation described in Ref. 24. The fits are made using the expression $\Delta H = \Delta H_0 + (\alpha/\gamma)f$ with $\Delta H_0 = 0$. Damping for both crystallographic directions [001] and [011] is practically the same $\alpha_{IrMn_3/Py} = (3.94 \pm 0.02) \times 10^{-2}$ but damping increased significantly in comparison to a simple layer of $Ni_{80}Fe_{20}$ (Py) $\alpha_{Py} = (8.66 \pm 0.03) \times 10^{-3}$, representing an increase of $\Delta\alpha = (3.06 \pm 0.02) \times 10^{-2}$ in the damping. Considering the variation of damping it is possible to get the spin-mixing conductance $g_{eff}^{\uparrow\downarrow} = 4\pi M_{eff} t_{FM} \Delta\alpha/(\gamma\hbar) = (1.6 \pm 0.1) \times 10^{16}$ cm$^{-2}$ [19], where $t_{FM} = 10$ nm is the thickness of the ferromagnetic material ($Ni_{80}Fe_{20}$, Py). This means that the interfacial coupling of the material modifies its dynamic properties.

We have also performed ferromagnetic resonance experiments in the presence of a *dc* current [See **Fig. 2 (a)**]. The experimental results suggest that there are three distinct contributions to $\alpha$: the first mechanism is that $\alpha$ is strongly facet-dependent and is derived from the antiferromagnetic domains of the uncompensated spins; the second mechanism is that it is facet-independent and arises from bulk spin-orbit coupling within the $IrMn_3$ layer and the third is the magnetic spin Hall effect (MSHE). It is known that chemically ordered $IrMn_3$ has a triangular chiral magnetic structure with the Mn magnetic moments aligned at 120° to each other in the (111) plane [20]. The coupling of the magnetization of $Ni_{80}Fe_{20}$ to the Mn interface moments becomes strongly fixed in its preferred direction by the preferred antiferromagnetic domains in the bulk of the $IrMn_3$ film [26-28]. The coupling at the $IrMn_3/Ni_{80}Fe_{20}$ interface is responsible for the spin current flow and the concomitant manipulation of damping of the magnetization in $Ni_{80}Fe_{20}$. **Figures 3 (a)** and **(c)** show the linewidth variation as a function of the resonance frequency for the magnetic field applied along the [001] and [011] crystallographic directions, respectively. Based on the Landau-Lifshitz-Gilbert-Slonczewski equation the linewidth as a function of the frequency can be written as $\Delta H = \Delta H_0 + (\alpha/\gamma)f + \theta_{MSH}^{eff}/(M_{eff}t_{FM})(\hbar/2e)j_c\mathscr{L}$ where $\theta_{MSH}^{eff}$ is the effective magnetic spin Hall angle, $e$ is the charge of electron, $j_c$ is the charge density and the spin transparency of interface is represented by $\mathscr{L} = [g_{eff}^{\uparrow\downarrow}\tan(L/2\lambda_{AF})]/[\sigma h/(2\lambda_{AF}e^2) + g_{eff}^{\uparrow\downarrow}\coth(L/\lambda_{AF})]$ [29], where $\lambda_{AF}$ is the spin diffusion length in the $IrMn_3$, $g_{eff}^{\uparrow\downarrow}$ is the effective spin mixing conductance, $\sigma$ is the electrical conductivity and $L$ is the thickness. For analyzing the measurements of the $IrMn_3$(20 nm)/Py(10 nm) samples with the magnetic field applied along the [001] direction we used the following parameters: resonance frequency $f = (17.45 \pm 0.03)$ GHz, linewidth of $\Delta H = (336.5 \pm 0.5)$ Oe, $dc$ current density of $j_c = 2.5\times 10^4$ Acm$^{-2}$, the spin diffusion length $\lambda_{AF} = 1$ nm [20], thickness $L = 20$ nm, and electrical conductivity $\sigma_{[001]}^{Elec} = (8.2 \pm 0.2)\times 10^3$ $\Omega^{-1}$cm$^{-1}$, which is in accordance with the previously reported values [20-22]; using these parameters we obtain an effective magnetic spin Hall angle of $\theta_{MSH}^{eff} \sim (0.33 \pm 0.02)$. This result is consistent with Ref. 20, where the spin Hall effect was probed via other techniques.

We observe that for the magnetic field applied along the [001] direction a change in the magnetic damping occurs, depending on the direction (positive or negative) of the applied $dc$ current. This does not occur when the magnetic field is applied along the [011] direction, which indicates a change of the current induced toques in $Ni_{80}Fe_{20}$ as a function of the magnetic field due to the MSHE of $IrMn_3$. Also note, that the electrical conductivity of $Ni_{80}Fe_{20}$ is $\sigma_{Py} = 2\times10^4$ $\Omega^{-1}$cm$^{-1}$ [30] and thus $\sigma_{Py}/\sigma_{[001]}^{Elec} = (2.4 \pm 1)$. Together with the thickness ratio of the two layers, this means that the $dc$ current flows in about equal parts through the $Ni_{80}Fe_{20}$ and $IrMn_3$ layers. At the same time $Ni_{80}Fe_{20}$ has a small spin Hall angle, $\theta_{SH} = 0.005$ [31] compared to $IrMn_3$ (*i.e.*, $\theta_{SH}/\theta_{MSH}^{eff} \approx 1.5\%$) and thus any magnetic torques originating from electric current flowing through the $Ni_{80}Fe_{20}$ layer can be neglected. **Figures 3 (b)** and **(d)** show the variation of the magnetic damping of the $Ni_{80}Fe_{20}$ layer as a function of the applied $dc$ current along the crystallographic [001] direction and the magnetic field applied in crystallographic [001] and [011] directions, respectively.

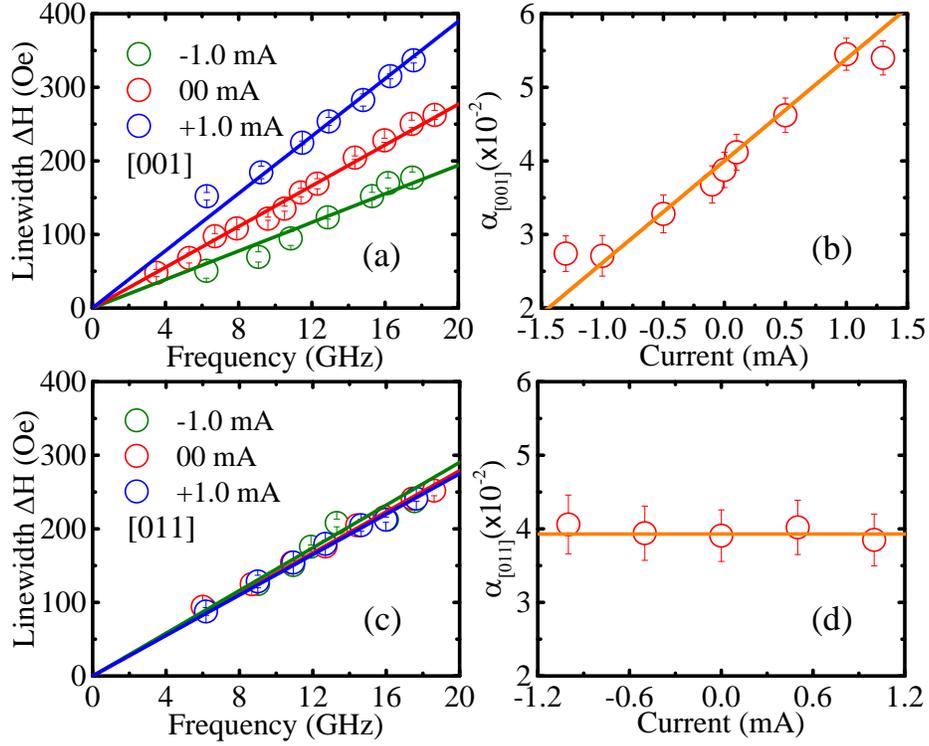

**Figure 3:** Electric current modulation of FMR measured for MgO/IrMn$_3$(20 nm)/Ni$_{80}$Fe$_{20}$(10 nm)/Ti(2 nm). **(a)** and **(c)** FMR linewidth variation as a function of the resonance frequency with *dc* currents of ±1 mA applied along the [001] crystallographic direction and magnetic field applied along the [001] and [011] crystallographic directions, respectively. **(b)** and **(d)** damping variation as a function of *dc* current for the [001] and [011] crystallographic directions, respectively.

The MSHE generated in IrMn$_3$ leads to significant changes in the magnetization dynamics of the adjacent Ni$_{80}$Fe$_{20}$ layer depending on the crystalline orientations. For an (100)-oriented IrMn$_3$ film, the in-plane current leads to a large out-of-plane spin current whose amplitude is much larger than that of a (111)-oriented IrMn$_3$ films [20]. As shown in **Fig. 1 (a)** in the face-centered cubic lattice of IrMn$_3$, the Mn atoms are arranged in the form of triangles within the {111} planes of the primitive crystallographic unit cell. Because of the inherent frustration of such triangular antiferromagnetic arrangements, neighboring Mn moments align non-collinear, at an angle of 120º to each other, to form two distinct antiferromagnetic arrangements, in which the Mn moments point either toward (Lat. 1) or away from the center of the triangle (Lat. 2), respectively [see **Fig. 4 (a)**]. Using either a mirror reflection or time-reversal operation, Lat. 1 and Lat. 2 can be transformed into each other. For the (011)-crystallographic plane, the reversal operation process should apply both to the lattice and magnetic moments and thus time-reversal only reverse directions of all moments, because either the mirror symmetry or the time-reversal symmetry is broken in IrMn$_3$. On the other hand, Lat. 1 and Lat. 2 are nonequivalent ground states and are chiral images of each other. In this case, Lat. 1 and Lat. 2 exhibits the same energies and both exist spontaneously in the material. It is known that under time reversal the magnetic spin Hall effect is odd [16] whereas the conventional spin Hall effect is even [31]. Thus, we can conclude with the help of time reversal that Lat. 1 and Lat. 2 will exhibit the same MSHE, as schematically shown in **Fig. 4**.

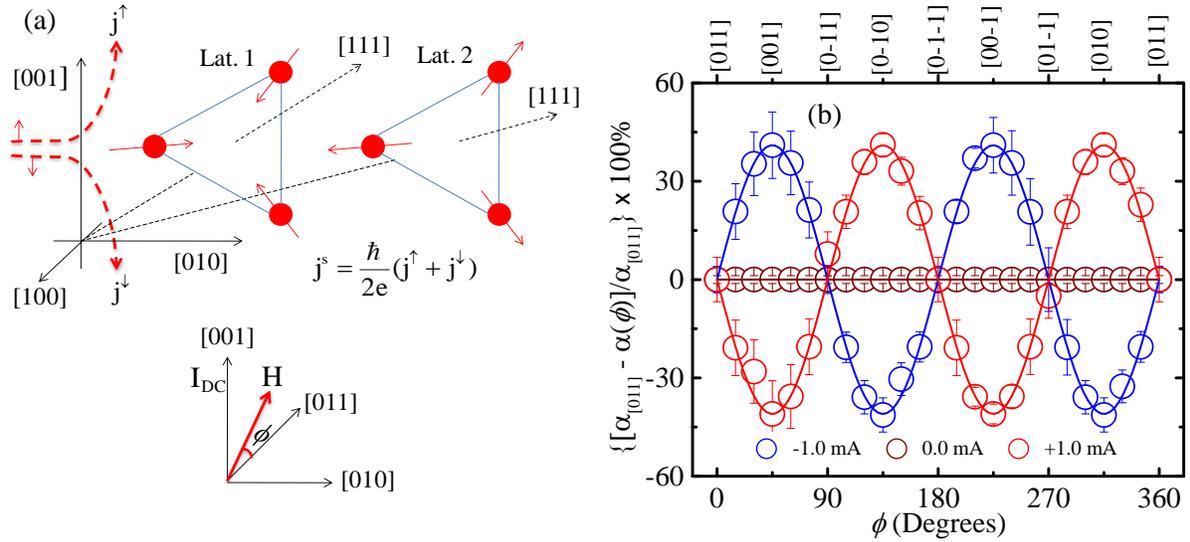

**Figure 4: (a)** Schematic diagram of two chiral antiferromagnetic lattices (Lat. 1 and Lat. 2) of the Mn moments inIrMn$_3$. Due to the magnetic spin Hall effect, the trajectories of spin-up and spin-down electrons have opposite transverse components and produce a spin current in the [001] direction. The angular variation of the magnetic field (H) in the plane of the film is also shown and it determines the polarization of the spin accumulation due to magnetic spin Hall effect. **(b)** Variation of the current dependent Gilbert magnetic damping as a function of the applied magnetic field angle, which in turn defines the polarization of the spin accumulation due to magnetic spin Hall effect. The dotted curve represents the function $\{[\alpha_{[011]} - \alpha(\phi)]/\alpha_{[011]}\}\times 100\% = \Delta\alpha_{MSHE}\sin(\pm 2\phi)$, with $\Delta\alpha_{MSHE} = (40 \pm 1)$ %.

**Figure 4 (b)** shows for a sample of MgO/IrMn$_3$(20nm)/Ni$_{80}$Fe$_{20}$(10nm)/Ti(2nm) the damping variation as a function of the applied magnetic field angle, which in turn defines the polarization of the spin accumulation due to the magnetic spin Hall effect. The modulation of the electric current induced damping modulation of the Ni$_{80}$Fe$_{20}$ as a function of the applied magnetic field direction provides an efficient pathway for modulating magnetization dynamics. The influence of the spin accumulation generated by the spin current $j^s = (j^\uparrow + j^\downarrow)\hbar/2e$ at the antiferromagnetic/ferromagnetic interface due to the MSHE also depends on the crystallographic direction of the material. As shown in **Fig. 3 (a)** for -1 mA or +1 mA *dc* current, the damping variation for magnetic fields applied either along the [001] or [011] crystallographic directions is $\Delta\alpha_{MSHE} \approx (40 \pm 1)$ %, which is consistent with the ratio between the intrinsic Hall conductivities [20-22]. The MSHE is influenced by the contributions of intrinsic Hall conductivities of the domains with opposite chirality of spin [32]. A variation of this order opens new possibilities for the control of spin currents and thus information flow in spintronics devices. The fit of **Fig. 4 (b)** was realized with the function $\{[\alpha_{[011]} - \alpha(\phi)]/\alpha_{[011]}\} \times 100\% = \Delta\alpha_{MSHE}\sin(\pm 2\phi)$, where $\Delta\alpha_{MSHE} = (40 \pm 1)$ % is consistent with the previous results. This signifies that the condition of mirror symmetry is broken in IrMn$_3$, providing evidence for the MSHE existing in this material [16].

Similar measurements were performed with *dc* electrical currents applied along the [011] crystallographic direction. The results are similar to the ones shown in **Figure 4 (b)**. It is also worthwhile to note that recently very large anisotropies of the magnetization damping as a function of magnetic field direction with respect to the crystalline orientation have been observed for individual ferromagnetic layers [33]. This differs from the current observation where the magnetic damping of the Ni$_{80}$Fe$_{20}$ film without any applied electric current is largely independent of the magnetic field orientation as can be seen in **Figs. 3 (b)** and **(d)**. In contrast here, only the electric current dependent part of the magnetization damping shows a large anisotropy.

In summary, we have shown that bilayers of IrMn$_3$/Ni$_{80}$Fe$_{20}$ have a strong modulation of electric current induced damping-like spin torques. The angular magnetic field dependence indicates that this damping-like torques originates from magnetic spin Hall effects in the IrMn$_3$. This indicates that chiral antiferromagnetic systems, such as IrMn$_3$ can provide additional functionality for electric current control of magnetization dynamics with very different symmetries than what can be expected from conventional spin Hall effects. Thus this work provides new perspectives for the fundamental understanding of charge- to spin-current conversions in antiferromagnets, as well as new avenues for integrating chiral antiferromagnets into spintronics devices.

**Acknowledgments**

We acknowledge useful discussion with Roland Winkler. All experimental work was performed at the Argonne National Laboratory and supported by the Department of Energy, Office of Science, Materials Science and Engineering Division. The use of the Centre for Nanoscale Materials was supported by the U. S. Department of Energy (DOE), Office of Sciences, Basic Energy Sciences (BES), under Contract No. DE-AC02-06CH11357. Part of the data analysis was also supported by the NSF through the University of Illinois at Urbana-Champaign Materials Research Science and Engineering Center DMR-1720633. José Holanda acknowledges the financial support of Conselho Nacional de Desenvolvimento Científico e Tecnológico (CNPq)-Brasil. Vedat Karakas and Ozhan Ozatay acknowledge financial support from TUBITAK grant 118F116 and Bogazici University Research Fund grant 17B03D3. Zhizhi Zhang acknowledges additional financial support from the China Scholarship Council (no. 201706160146) for a research stay at Argonne.